\def\papertitle{CONMOD: Controllable Neural Frame-based Modulation Effects}
\def\paperauthorA{Gyubin Lee}
\def\paperauthorB{Hounsu Kim}
\def\paperauthorC{Junwon Lee}
\def\paperauthorD{Juhan Nam}

\documentclass[twoside,a4paper]{article}
\usepackage{etoolbox}
\usepackage{dafx_24}
\usepackage{amsmath,amssymb,amsfonts,amsthm}
\usepackage{euscript}
\usepackage[T1]{fontenc}
\usepackage[utf8]{inputenc}
\usepackage{ifpdf}
\usepackage[english]{babel}
\usepackage{caption}
\usepackage{subfig} 
\usepackage{color}
\usepackage{makecell, multirow}
\usepackage{xurl}
\makeatletter
\newcommand*{\rom}[1]{\expandafter\@slowromancap\romannumeral #1@}
\makeatother

\input glyphtounicode
\ninept

\newcounter{numauth}
\newcounter{listcnt}
\newcommand\authcnt[1]{\ifdefined#1 \stepcounter{numauth} \fi}

\newcommand\addauth[1]{
\ifdefined#1 
\stepcounter{listcnt}
\ifnum \value{listcnt}<\value{numauth}
\appto\authorslist{, #1}
\else
\appto\authorslist{~and~#1}
\fi
\fi}
\authcnt{\paperauthorB}
\authcnt{\paperauthorC}
\authcnt{\paperauthorD}
\authcnt{\paperauthorE}
\authcnt{\paperauthorF}
\authcnt{\paperauthorG}
\authcnt{\paperauthorH}
\authcnt{\paperauthorI}
\authcnt{\paperauthorJ}
\def\authorslist{\paperauthorA}
\addauth{\paperauthorB}
\addauth{\paperauthorC}
\addauth{\paperauthorD}
\addauth{\paperauthorE}
\addauth{\paperauthorF}
\addauth{\paperauthorG}
\addauth{\paperauthorH}
\addauth{\paperauthorI}
\addauth{\paperauthorJ}

\usepackage{times}

\newif\ifpdf
\ifx\pdfoutput\relax
\else
   \ifcase\pdfoutput
      \pdffalse
   \else
      \pdftrue
\fi

\ifpdf 
  \usepackage[pdftex,
    pdftitle={\papertitle},
    pdfauthor={\authorslist},
    pdfsubject={Proceedings of the 27th International Conference on Digital Audio Effects (DAFx24)},
    colorlinks=false, 
    bookmarksnumbered, 
    pdfstartview=XYZ 
  ]{hyperref}
  \usepackage[pdftex]{graphicx}
\else 
  \usepackage[dvips]{epsfig,graphicx}
  \usepackage[dvips,
    pdftitle={\papertitle},
    pdfauthor={\authorslist},
    pdfsubject={Proceedings of the 27th International Conference on Digital Audio Effects (DAFx24)},
    colorlinks=false, 
    bookmarksnumbered, 
    pdfstartview=XYZ 
  ]{hyperref}
\fi
\usepackage[hypcap=true]{caption}
\title{\papertitle}

\threeaffiliations{
\paperauthorA \,\thanks{\vspace{-3mm}}}
{\href{https://cs.kaist.ac.kr}{School of Computing} \\ KAIST \\ Daejeon, South Korea\\
{\tt \href{mailto:gbstorm81@kaist.ac.kr}{gbstorm81@kaist.ac.kr}}
}
{\paperauthorB \,}
{\href{https://ct.kaist.ac.kr}{Graduate School of Culture Technology} \\ KAIST \\ Daejeon, South Korea\\ {\tt \href{mailto:hanshounsu@kaist.ac.kr}{hanshounsu@kaist.ac.kr}}
}
{\paperauthorC \,}
{\href{https://ct.kaist.ac.kr}{Graduate School of Culture Technology} \\ KAIST \\ Daejeon, South Korea\\
{\tt \href{mailto:juhan.nam@kaist.ac.kr}{juhan.nam@kaist.ac.kr}}
}
\fouraffiliations{
\paperauthorA \,\thanks{\vspace{-3mm}}}
{\href{https://cs.kaist.ac.kr}{School of Computing} \\ KAIST \\ Daejeon, South Korea\\
{\tt \href{mailto:gbstorm81@kaist.ac.kr}{gbstorm81@kaist.ac.kr}}
}
{\paperauthorB \,}
{\href{https://ct.kaist.ac.kr}{Graduate School of Culture Technology} \\ KAIST \\ Daejeon, South Korea\\ {\tt \href{mailto:hanshounsu@kaist.ac.kr}{hanshounsu@kaist.ac.kr}}
}
{\paperauthorC \,}
{\href{https://gsai.kaist.ac.kr}{Kim Jaechul Graduate School of AI} \\ KAIST \\ Daejeon, South Korea\\
{\tt \href{mailto:james39@kaist.ac.kr}{james39@kaist.ac.kr}}
}
{\paperauthorD \,}
{\href{https://ct.kaist.ac.kr}{Graduate School of Culture Technology} \\ KAIST \\ Daejeon, South Korea\\
{\tt \href{mailto:juhan.nam@kaist.ac.kr}{juhan.nam@kaist.ac.kr}}
}

\begin{document}
\ifpdf 
  \DeclareGraphicsExtensions{.png,.jpg,.pdf}
\else  
  \DeclareGraphicsExtensions{.eps}
\fi


\maketitle

\begin{abstract}
Deep learning models have seen widespread use in modelling LFO-driven audio effects, such as phaser and flanger. Although existing neural architectures exhibit high-quality emulation of individual effects, they do not possess the capability to manipulate the output via control parameters. To address this issue, we introduce \textbf{Co}ntrollable \textbf{N}eural Frame-based \textbf{Mod}ulation Effects (CONMOD), a single black-box model which emulates various LFO-driven effects in a frame-wise manner, offering control over LFO frequency and feedback parameters. Additionally, the model is capable of learning the continuous embedding space of two distinct phaser effects, enabling us to steer between effects and achieve creative outputs. Our model outperforms previous work while possessing both controllability and universality, presenting opportunities to enhance creativity in modern LFO-driven audio effects. Additional demo of our model is available in the accompanying website.\footnote{\url{https://lgbin81.notion.site/Audio-Demo-for-DAFx-2024-Paper-70705a992d5c4730a1f70fbaf951243b?pvs=4}}
\end{abstract}

\section{Introduction}
\label{sec:intro}

The study of virtual analog (VA) modelling via deep neural networks have become a well-established topic nowadays. Neural networks are strongly capable of modelling the behavior of the system in a black-box manner, and is being applied in various VA modelling tasks \cite{vanhatalo2022reviewamp, simionato2023fully, comunita2023modelling, carson2023differentiable}. Furthermore, the strongpoint of deep neural networks as a regression model anticipates an intuitive and creative effect by adjusting a representative embedding vector of the trained model \cite{kim2019neural, steinmetz2021steerable}.

The field of VA is largely divided into white-box, grey-box and black-box approaches. The white-box approach discretizes the acoustic system of the analog circuit structure to obtain a digital model \cite{eichas2014phaserwhitebox, mavcak2016flangerwhitebox}. The black-box approach incorporates parametric models which emulate the input-output behavior of the analog circuits without knowing the physical structure \cite{helie2009volterra, eichas2016black, ramirez2019general}. Grey-box approach leverages some knowledge of the acoustic system to better emulate the parametric model \cite{kiiski2016time, colonel2022reverse}. 
Among them, the black-box approach has recently drawn a great attention as it can behave more robust and general in the case of deep neural networks when the model is well-structured and trained with a sufficient amount of data \cite{martinez2018equalization, ramirez2019general, martinez2020deep, steinmetz2022efficient, simionato2023fully, comunita2023modelling}.

Phaser and flanger effects modulate the input audio using a periodic signal typically within the human preferred range (0.1-10Hz), commonly referred to as LFO-driven effects. Core aspect of modelling these effects is maintaining precise periodicity regardless of the input audio duration. Therefore, early neural architectures integrated an LFO signal in the form of a rectified sine wave \cite{wright2020neural, wright2021neuralmodeling} or a sinusoidal wave \cite{carson2023differentiable}  to enable the model to focus on generating the spectral structure. Early works related to these approach \cite{wright2020neural, wright2021neuralmodeling} extracts the LFO signal manually, which is a clear disadvantage for training. \cite{mitcheltree2023modulation} further proposed a CNN-based LFO extraction model, achieving plausible results for unseen audio and effects. Unlike existing black-box approaches that learn the direct mapping from dry input audio sets to corresponding wet output audio sets, Carson et al. simplifies the role of model into emulating solely the effect of a phaser pedal using a frame-based processing \cite{carson2023differentiable}. Nevertheless, the paper does not explore the model's controllability with respect to control parameters(e.g. LFO frequency, feedback, etc), even though directly manipulating the sinusoidal wave may alter the LFO frequency. And even so, the model is controllable with only a single control parameter.

In this paper, we build upon the model architecture proposed by Carson et al. \cite{carson2023differentiable} and introduce a novel controllable neural effects designed for two types of modulation effects, phaser and flanger. Specifically, our neural architecture enables adjustment of the output with user-specified rate and feedback parameters for the target effect. Rather than directly optimizing the parameters related to the physics of the phaser effect, we optimize a neural architecture consisting of long-term short-memory (LSTM), multi-layer perceptron (MLP), and conditional feature wise linear modulation (FiLM) blocks\cite{perez2018film}. To enhance the model's controllability on the LFO frequency, we introduce a training procedure which optimizes multiple LFO signals within a single neural network. We compare our work with a baseline model \cite{carson2023differentiable} and show that our model shows superior accuracy in trained control parameter settings, while also exhibiting impressive performance with unseen control parameters. Inspired by \cite{kim2019neural, steinmetz2021steerable}, we further explored our model's capability to train multiple modulation effects simultaneously. Through training on two unique phaser effects concurrently, and by manipulating each effect's learned embedding, we show that our model can produce intriguing mixed results between the two distinct phaser effects. The main contributions of our work can be summarized as follows:
\begin{itemize}
    \item We introduce a controllable neural frame-based modulation effect pedal targeting phaser and flanger, which outperforms previous model in emulation capability.
    \item Our model possesses robust controllability on multiple control parameters, capable of modelling effects on unseen control parameter settings, a feat unattainable by previous models.
    \item For the first time, we train a single model on two distinct phaser effects and demonstrate the model's steerability by manipulating the learned representative embeddings.
\end{itemize}



\section{Target Modulation Effects}

In this section, we briefly explain the characteristics of our target modulation effects, and the specific effect pedals that were trained on our model. For each of the effects, we trained our model with digital and analog devices.

\subsection{Phaser effects and its controls}

The phaser pedal modulates the input signal by utilizing low frequency oscillating notch filters or low-order all-pass filters. When cascaded, these filters periodically attenuate certain frequency levels of the signal, thereby introducing soft, moving sounds. In this paper, we focus on learning to control the two main parameters: LFO frequency and feedback, following \cite{carson2023differentiable}. LFO frequency, also called \textit{Speed} or \textit{Rate}, governs the frequency of the LFO signal, while feedback adjusts the amount of filtered output signal that is fed back to the input. The adjustment of feedback leads to an overall change in the shape of the frequency response, notably bringing resonance peaks in the unattenuated area \cite{beigel1979digital}. Feedback is often termed \textit{Color} due to its ability to alter the frequency response shape.

\subsubsection{Studio One Phaser$^2$}

Studio One Phaser$^2$ is a built-in digital phaser included in Studio One, featuring various controllable parameters such as Speed, Feedback, Center, and Range \cite{s1manualpresonus}. In Phaser$^2$ the \textit{Speed} knob, which controls the LFO frequency, allows for continuous adjustment within a range of 0 Hz to 10 Hz. The \textit{Feedback} knob also offers continuous range from 0 to 95\%. For the remaining controls, we maintain the default values.

\subsubsection{Moore Liquid Phaser}

Moore Liquid Phaser is a modern analog phaser offering selection among 5 distinct \textit{Phaser Type}s and featuring controls among 3 different \textit{Waveform Type}s(Round, Triangular, or Square), \textit{Speed}, and \textit{Color}. We set the \textit{Waveform Type} to Round, and the \textit{Phaser Type} to feedback mode, which corresponds to the 12 o' clock direction of the knob ('\rom{3}').

\subsection{Flanger effects and its controls}

The flanger effect has a similar effect compared to the phaser effect, generating spectral notches. The main difference is that flanger creates notches by introducing a variable-length feedforward delayline which leads to a comb filter. Therefore the notches, in general, are regularly distributed across the frequency spectrum, extending infinitely. In this work, we adopt a similar approach as in the phaser case, concentrating on the LFO frequency and feedback parameters, which function on the same principle as in the phaser effect.

\subsubsection{Studio One Flanger}

Studio One Flanger is a built-in digital flanger with controllable parameters similar to those found in a phaser, such as \textit{Speed}, \textit{Feedback}, \textit{Delay}, and \textit{LFO amount}. The \textit{Speed} control allows for adjustments within a continuous value from 0.01Hz to 10Hz. The \textit{Feedback} knob provides a range from -90\% to 90\%, where negative values lead to inverted feedback \cite{s1manualpresonus}.


\subsubsection{Flamma FC15 Analog Flanger}

The Flamma FC 15 analog flanger is a classic device carrying two effect modes, namely filter and normal, along with controllable knobs for \textit{Color}, \textit{Range}, and \textit{Rate} \cite{flammainnovation}. We maintain the \textit{Modes} setting at normal, and set the \textit{Range} parameter to the 12 o' clock position, while training our model on variable settings for \textit{Color} and \textit{Rate}.

\section{Methodology}
We introduce a novel black-box methodology controllable via various effect parameters, which we refer to as CONMOD(\textbf{Co}ntrollable \textbf{N}eural Frame-based \textbf{Mod}ulation Effects). Our model efficiently models the audio effect frame-wise with less reliance on prior domain knowledge.

\subsection{Model}\label{ssec:train}
Our model predicts the frame-dependent transfer function of the target effect using a sinusoidal LFO signal with trainable frequency and initial phase as input, following \cite{carson2023differentiable}. 
This approach approximates the effects, which are analogous to an IIR filter, by representing them as a truncated impulse response, where the truncation length is equal to the window size of each frame.
As the frame-based process remains differentiable throughout the system, we can train the model to predict solely the modulation effect, with the training objective being the target wet audio waveform. The input for the training phase is a chirp signal, and the target pair is the wet chirp signal.

\begin{figure}[t!]
\centerline{\includegraphics[width=\columnwidth]{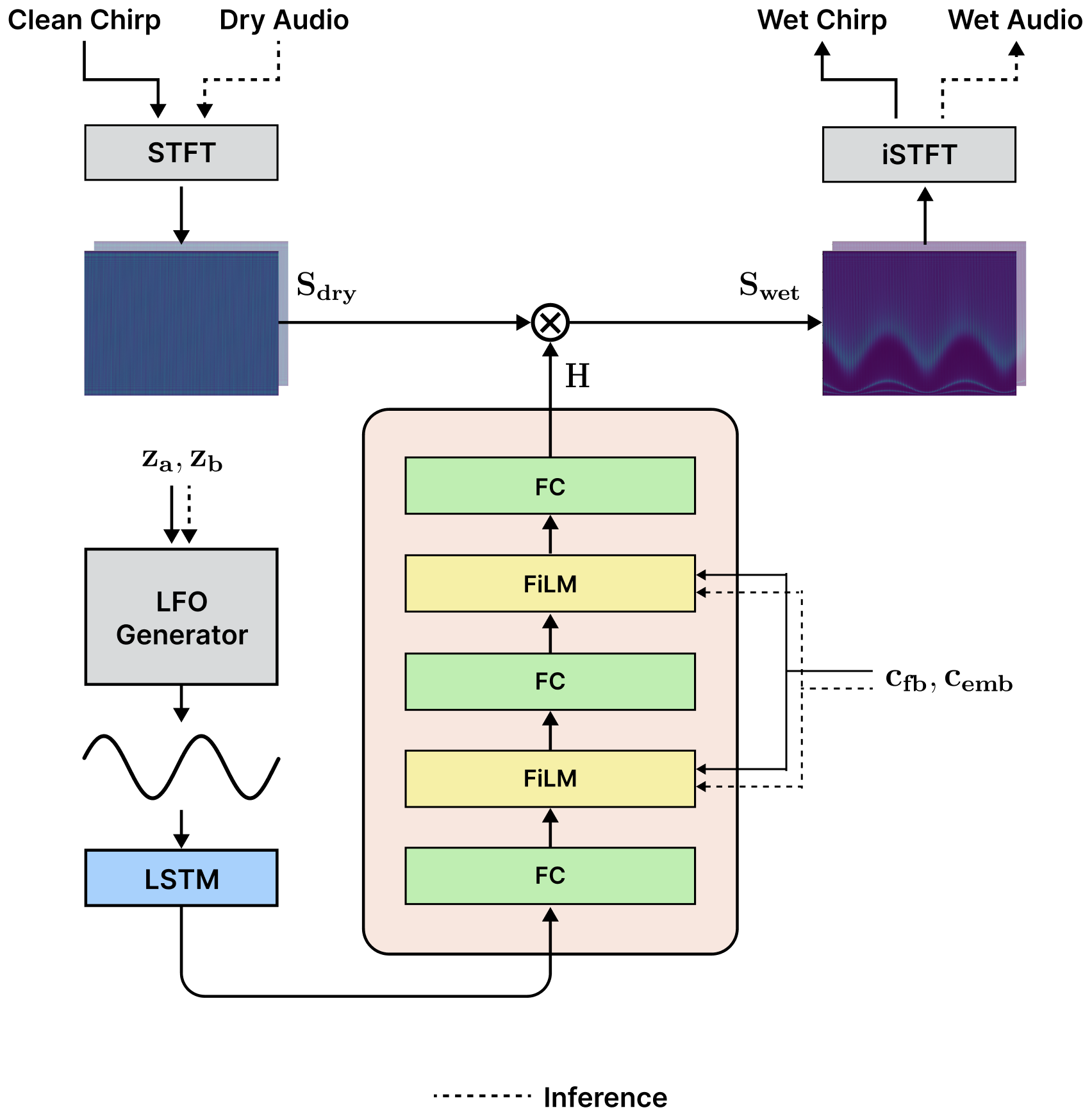}}
\caption{Overall model architecture. Blocks that have both dotted lines and solid lines as input indicate that the input data is distinctive for the training phase and inference phase. Grey-colored blocks indicate they have no trainable parameters and their processes are differentiable, thus enabling backpropagation. $\mathbf{z_a}$, $\mathbf{z_b}$, and $\mathbf{c_{emb}}$ are trainable parameters.}
\label{fig:model architecture}
\end{figure}

The main difference with \cite{carson2023differentiable} lies in the neural network structure responsible for predicting the transfer function. We begin by employing LSTM on the LFO signal to enhance the model's robustness under varying LFO frequency conditions. The features are then passed through fully-connected (FC) layers to shape the frame-dependent transfer function. FiLM layers are inserted between these FC layers to incorporate the feedback condition and effect type embeddings. The final output comprises predicted transfer functions for each input frame, resulting in a complex matrix equivalent in size to the STFT matrix of the dry signal. After element-wise multiplication of these matrices, an inverse STFT is conducted to derive the predicted wet signal. The comprehensive model architecture is depicted in Figure \ref{fig:model architecture}.

Our model employs both time domain loss and frequency domain loss for the training objective. For the time domain loss, we utilize the error-to-signal ratio (ESR) defined as:
\begin{equation}
    \mathcal{L}_{ESR}(y, \hat{y}) = \frac{\sum_{l=0}^{L-1}(y[l]-\hat{y}[l])^2}{\sum_{l=0}^{L-1}y[l]^2}
\end{equation}
where $L$ is the length of the training data in samples. For the frequency domain loss, we adopt the multi-resolution spectral loss (MRSL) defined as:
\begin{equation}
    \mathcal{L}_{MRSL} = \sum_{i} ||S_i - \hat{S_i}||_1 + ||\log S_i - \log\hat{S_i}||_1
\end{equation}
where $i$ is the index of each FFT size, which is set as (512, 1024, 2048). The neighboring frames during the STFT are overlapped by 75\%. Consequently, the final training objective of our model becomes: 
\begin{equation}
    \mathcal{L}(y, \hat{y}) = \lambda \mathcal{L}_{ESR}(y, \hat{y})+\mathcal{L}_{MRSL}
\end{equation}
where the weight $\lambda$ is set to 2 throughout all experiments.


\subsection{Conditioning the control parameters}

We employed the FiLM layer to apply the feedback condition to the model \cite{perez2018film}. This is mathematically formalized as the following:
\begin{equation}
\begin{split}
    & \text{MLP}([\mathbf{c_{fb}}, \mathbf{c_{emb}}]) = [\gamma, \beta] \\
    & \text{FiLM:\ \ } \mathbf{y_{fc}} = \gamma \cdot \mathbf{x_{fc}} + \beta
\end{split}
\end{equation}
where $\mathbf{c_{fb}}$ denotes the feedback, $\mathbf{c_{emb}}$ is the embedding of effect type, $[\cdot,\cdot]$ denotes simple concatenation of two vectors, $\gamma$, $\beta$ are the output factors of FiLM, and $\mathbf{x_{fc}}$, $\mathbf{y_{fc}}$ are the intermediate input feature and its affine transformed output respectively.\def\hide{We show that this simple yet powerful approach effectively controls the effector sound.} Effect type embedding $\mathbf{c_{emb}}$ is only used when the model is trained on multiple phaser effects. (refer to Section \ref{sec:multi_phaser}).



\begin{figure}[t]
\centerline{\includegraphics[width=7.5cm]{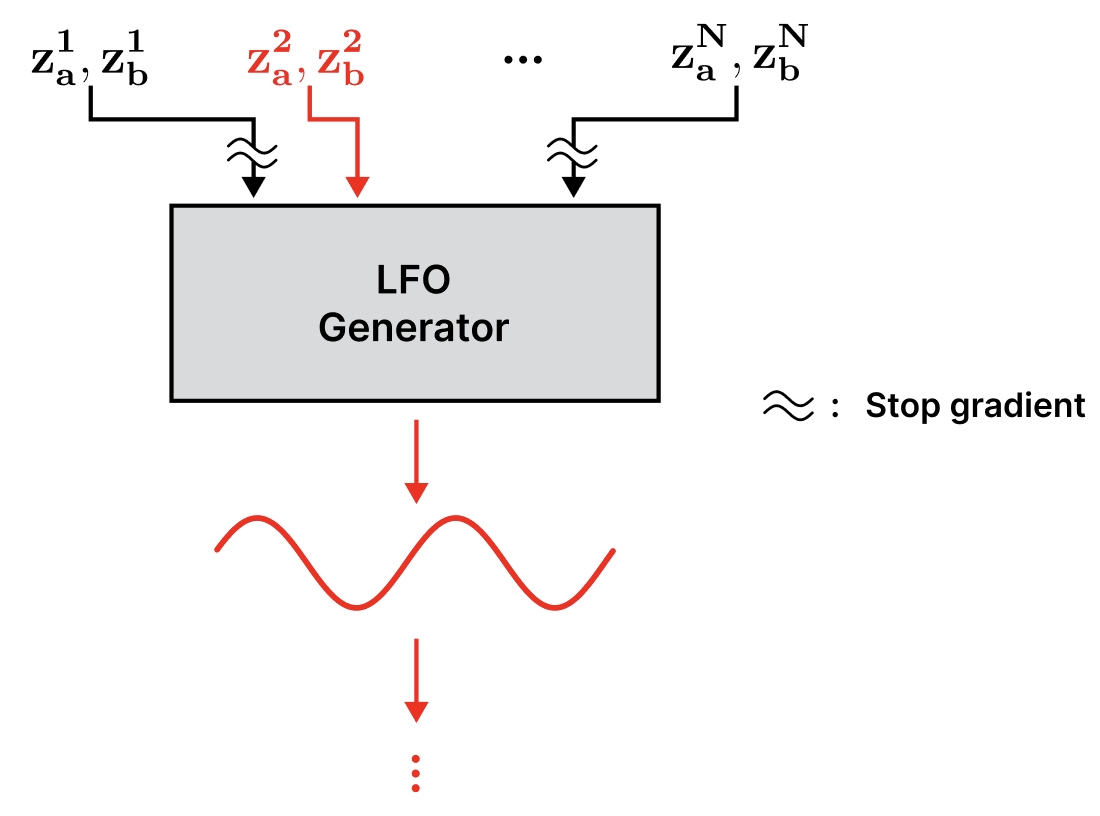}}
\caption{Proposed model training technique on multiple LFO frequency controls. For a random input-output pair with specified LFO frequency settings, only the corresponding $z_a$, $z_b$ parameters are optimized. Figure depicts the case when $z_a^2$, $z_b^2$ are optimized. Other parameters are not updated by employing stop gradient operation.}
\label{fig:multi LFO training}
\end{figure}

In \cite{carson2023differentiable}, the time-varying behavior of the target wet audio is modeled as a sinusoidal LFO signal and is fitted through gradient descent during the training phase \cite{hayes2023sinusoidal}. The sinusoidal LFO signal is generated from input parameters directly related to the frequency $\mathbf{z_a}$ and initial phase $\mathbf{z_b}$, which is then fed into the LSTM block. In our approach, we follow the LFO fitting methodology in \cite{carson2023differentiable} and additionally train the model on target wet audio modulated with multiple LFO frequencies to achieve robust controllability over frequency. The parameters being optimized are thus $(z_a^1, z_b^1), (z_a^2, z_b^2)$, ..., $(z_a^N, z_b^N)$ where each of the $N$ couple of parameters is assigned to train different LFO signals. When an audio pair with LFO frequency corresponding to the $i$-th signal is input to the model, other parameters are not updated, utilizing the stop gradient operator \cite{van2017neural}, as illustrated in Figure \ref{fig:multi LFO training}.

\section{Experiments}

\subsection{Dataset Preparation}

We utilize a chirp-train signal with impulse period 40ms, passed through cascaded 64 all-pass filters with a coefficient of $p = 0.9$, following \cite{carson2023differentiable}. Training data consists of 6.67 seconds of chirp signal, while direct-input (DI) recordings provided in \cite{carson2023differentiable}\footnote{https://github.com/a-carson/ddsp-phaser/tree/main/audio\_data} are utilized for inference.
For digital phaser and flanger effects, we vary the LFO frequency(Hz) among $0.23, 0.73, 1.13$ and the feedback (\%) parameters among $0, 25, 50, 75$. The number of control parameter values are selected heuristically, resulting in a total of 12 audio samples used for training the model. All other effect parameters except for LFO frequency and feedback parameter remain fixed. Overall parameter settings are organized in Table \ref{tab:digital_parameter_settings}  Notably, 23, 73, and 113 are all prime numbers, ensuring that the LFO generator provides various and non-overlapping LFO phase to the LSTM block.

\begin{table}[h]
\caption{\itshape Control parameter experiment settings for the digital effect.}
\label{tab:digital_parameter_settings}
\centering
\begin{tabular}{ccc}
\Xhline{1.5pt}
\multicolumn{3}{c}{\textbf{Studio One Digital Phaser (Phaser$^2$)}} \\ \hline
\multicolumn{1}{c|}{\multirow{2}{*}{Modulation}} & \multicolumn{1}{c|}{Stages} & 6 \\
\multicolumn{1}{c|}{} & \multicolumn{1}{c|}{Speed (Hz)} & 0.23, 0.73, 1.13 \\ \hline
\multicolumn{1}{c|}{\multirow{3}{*}{Color}} & \multicolumn{1}{c|}{Center}        & 1300                     \\
\multicolumn{1}{c|}{} & \multicolumn{1}{c|}{Range} & 80\% \\
\multicolumn{1}{c|}{} & \multicolumn{1}{c|}{Feedback (\%)} & 0, 25, 50, 75 \\ \hline
\multicolumn{1}{c|}{\multirow{2}{*}{Global}} & \multicolumn{1}{c|}{Stereo Spread} & 2.0\% \\
\multicolumn{1}{c|}{} & \multicolumn{1}{c|}{Mix} & 70.0\% \\ 
\Xhline{1.5pt}
\multicolumn{3}{c}{\textbf{Studio One Digital Flanger}} \\ \hline
\multicolumn{1}{c|}{\multirow{2}{*}{Color}} & \multicolumn{1}{c|}{LFO-Amount} & 80\% \\
\multicolumn{1}{c|}{} & \multicolumn{1}{c|}{Speed (Hz)} & 0.23, 0.73, 1.13 \\ \hline
\multicolumn{1}{c|}{\multirow{2}{*}{Modulation}} & \multicolumn{1}{c|}{Feedback (\%)} & 0, 25, 50, 75 \\
\multicolumn{1}{c|}{} & \multicolumn{1}{c|}{Delay} & 1.50ms \\ \hline
\multicolumn{1}{c|}{Global} & \multicolumn{1}{c|}{Mix} & 70.0\% \\ \hline
\end{tabular}
\end{table}

The LFO frequency of the analog effect was adjusted to closely match the LFO settings of the digital effects. For the analog phaser, the LFO frequency knob was set to 8 o'clock, halfway between 8 o'clock and 9 o'clock, and 9 o'clock, resulting in corresponding LFO frequencies (in Hz) of ${0.3549, \allowbreak 0.7519, \allowbreak 1.430}$. Similarly, for the analog flanger, the LFO frequency knob was set to 12 o'clock, 3 o'clock, and halfway between 3 o'clock and 4 o'clock, resulting in corresponding LFO frequencies of ${0.2438, 0.6816, 1.004}$. Regarding the feedback parameter, we determined the maximum feedback value that produced an audible sound (3 o'clock for the analog phaser and 12 o'clock for the analog flanger). Subsequently, we set four feedback values that evenly divided the range between no feedback and the maximum audible feedback. We designated each feedback such that a clockwise rotation of the knob by ${45^\circ}$ corresponds to a 12.5 increment in feedback. Consequently, the resulting feedback labels are ${0.0, 12.5, 37.5, 62.5}$ for the analog phaser and ${0.0, 12.5, 25, 37.5}$ for the analog flanger. As with the digital effects, all other effect parameters except for the LFO frequency and feedback parameters, are held constant. Specific parameter settings are presented in Table \ref{tab:analog_parameter_settings}.

\begin{table}[h!]
\caption{\itshape Control parameter experiment settings for the analog effect.}
\label{tab:analog_parameter_settings}
\centering
\begin{tabular}{cc}
\Xhline{1.5pt}
\multicolumn{2}{c}{\textbf{Moore Liquid Phaser}} \\
\hline
\multicolumn{1}{c|}{Phaser Type}     & 12 o' clock     \\
\multicolumn{1}{c|}{LFO shape} & Round \\
\multicolumn{1}{c|}{Speed (Hz)}     & 0.3549, 0.7519, 1.430 \\
\multicolumn{1}{c|}{Color}     & 0.0, 12.5, 37.5, 62.5  \\
\hline
\Xhline{1.5pt}
\multicolumn{2}{c}{\textbf{Flamma FC15 Analog Flanger}} \\
\hline
\multicolumn{1}{c|}{Mode}      & Normal \\
\multicolumn{1}{c|}{Range}     & 12 o'clock \\ 
\multicolumn{1}{c|}{Rate (Hz)}      & 0.2438, 0.6816, 1.004 \\
\multicolumn{1}{c|}{Color}     & 0.0, 12.5, 25, 37.5   \\
\hline
\end{tabular}
\end{table}

\subsection{Baselines}
For model comparison, we choose the previous grey-box phaser model\cite{carson2023differentiable} as our baseline. This model has a frame-based approach similar to our proposed model but is limited in learning the phaser effect due to its confined structure. Consequently, it cannot be trained for other time-varying filters or delay-based audio effects such as flanger. \def\hide{Additionally, the baseline model lacks controllability, meaning it can only generate a single combination of parameters.} Additionally, although the baseline model may have the potential for controllability by manipulating the LFO signal, this capability is not explicitly explored in the paper. Therefore, multiple baseline models are trained individually for each LFO frequency and feedback condition. We conducted the comparison for both digital and analog phasers.

\subsection{Training and Inference Setup}
To ensure that all LFO-related parameters are learned independently, all models are trained with a batch size of 1. We train audio with sample rate of 44.1kHz and adopt an FFT size of 4096, hop size of 441 and a fixed frame size of 1764 throughout the experiments. Adam optimizer with exponential learning rate decay ($\gamma = 0.9997$) and an initial learning rate of 0.001 is employed. The models are trained for 10,000 epochs while each epoch is composed of 12 chirp signals, resulting in a total training duration of 80.04 seconds. The hidden size of LSTM layer and MLP layers are 32 and 512 respectively, where a single MLP layer predicts the real and imaginary part of the STFT matrix. With these settings, the model has total 2.4M learnable parameters. For the digital effect, the initial LFO phase is identical for each recording. Consequently, only three LFO generator parameters, corresponding to the variations in LFO frequency, need to be trained. In the case of analog effect, we could not designate a desired \def\hide{equalize the }initial LFO phase during the recording of the 12 wet targets. Therefore the model had to predict 12 distinct LFO signals for each of the 12 wet chirp signals, leading to unstable convergence. Since we cannot obtain the groundtruth initial phase and LFO frequency values for the analog effect, we conducted pilot experiments by training the 12 signals individually and provide the model the trained $\{z_a^i\}_{i=1}^{12}$, $\{z_b^i\}_{i=1}^{12}$ as initial values to stabilize the training.

For quantitative evaluation, we utilized the ESR metric on 10s direct-input(DI) guitar recording.


\section{Results}

In this section, we aim to evaluate the performance of our model in modelling digital and analog phaser/flanger effects compared to the baseline model. We also highlight our model's controllability across various control parameter settings.
Ablation studies on long-term audio inference capability and the effect of training on multiple LFO frequencies are conducted. Lastly, we showcase a model trained on two distinct phasers simultaneously, which implies the potential of our approach in universal modelling.

\subsection{Main Results}


\subsubsection{Digital Phaser}
Table \ref{tab:digital_phaser} presents the performance of our model targeting digital phaser under seen control parameter settings. Results show a consistent trend of increased ESR with higher feedback. \def\hide{This trend is due to the fact that higher feedback augments circuit complexity, thereby complicating the transfer function prediction.} As shown in the table, our model significantly outperforms the baseline model.

\begin{table}[h!]
\caption{\itshape ESR(\%) results for model trained on the digital phaser (Studio One Phaser$^2$) under seen control parameter conditions. Training separate models for each control parameter configuration obtained results for the baseline model\cite{carson2023differentiable}.}
\centering
\begin{tabular}{cccc}
\hline
\multirow{2}{*}{LFO freq.(Hz)} & \multirow{2}{*}{Feedback(\%)} & \multicolumn{2}{c}{ESR(\%)} \\ \cline{3-4} 
& & CONMOD & Baseline\cite{carson2023differentiable} \\ \Xhline{1.5pt}
\multirow{4}{*}{0.23} & 0  & 0.30 & 18.3 \\
& 25 & 0.21 & 19.6 \\
& 50 & 0.86 & 22.6 \\
& 75 & 4.34 & 32.2 \\ \hline
\multirow{4}{*}{0.73} & 0 & 0.08 & 19.1 \\
& 25 & 0.09 & 21.9 \\
& 50 & 0.21 & 23.7 \\
& 75 & 2.20 & 34.3 \\ \hline
\multirow{4}{*}{1.13} & 0 & 0.07 & 38.6 \\
& 25 & 0.08 & 22.4 \\
& 50 & 0.17 & 22.8 \\
& 75 & 1.96 & 49.8 \\ \hline
\end{tabular}
\label{tab:digital_phaser}
\end{table}

\begin{figure}[t!]
\centerline{\includegraphics[width=\columnwidth]{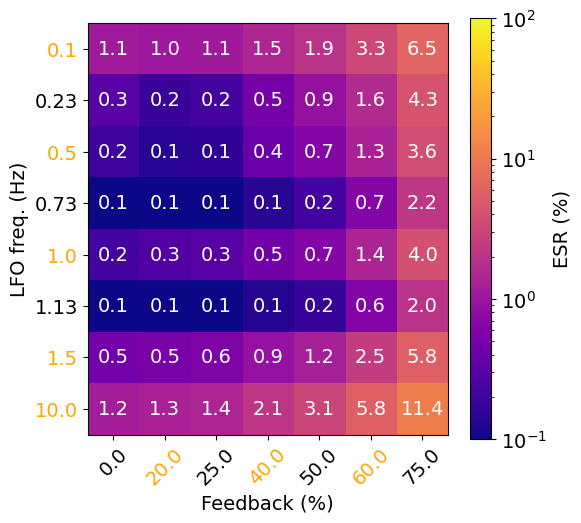}}
\caption{\itshape ESR(\%) results for model trained on the digital phaser under seen and unseen LFO frequencies and Feedback parameters. Seen cases are labeled black, and orange for unseen cases.}
\label{fig:digital_phaser}
\end{figure}

Figure \ref{fig:digital_phaser} shows the model inference score for both seen and unseen LFO frequencies and feedback during the training. 
Evaluations with unseen parameter configurations have been included to assess the model's controllability.
We found that the model demonstrates robust inference capabilities for unseen LFO frequencies, even outside the range of the training data. This indicates that the LSTM in our model is capable of predicting the spectral behavior for the varying LFO frequencies. We also found that our model shows consistent performance for unseen feedback conditions. 
It is evident that the FiLM layer in our model demonstrates strong performance in facilitating continuous control inferences.

\subsubsection{Analog Phaser}
\begin{figure}[ht]
\centerline{\includegraphics[width=7.5cm]{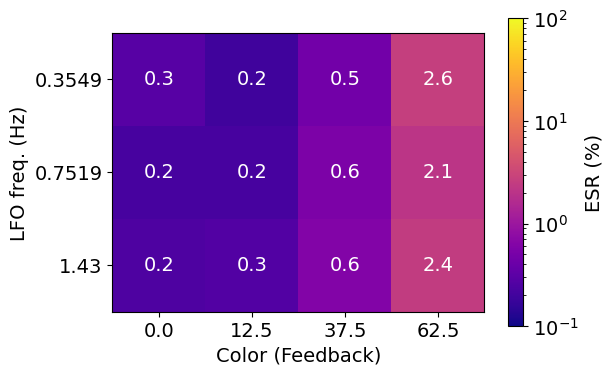}}
\caption{\itshape ESR(\%) results for model trained on the analog phaser under seen LFO frequencies and Color parameters.}
\label{fig:analog_phaser}
\end{figure}

Figure \ref{fig:analog_phaser} outlines the ESR results for the analog phaser model under seen control parameter settings. Similar to digital phaser, increased feedback leads to reduced accuracy. Notably, the baseline model failed to converge. We attribute this to the baseline structure learning model parameters assuming that the phaser is comprised of cascaded first-order all-pass filters.

\begin{figure}[ht]
\centerline{\includegraphics[width=7.5cm]{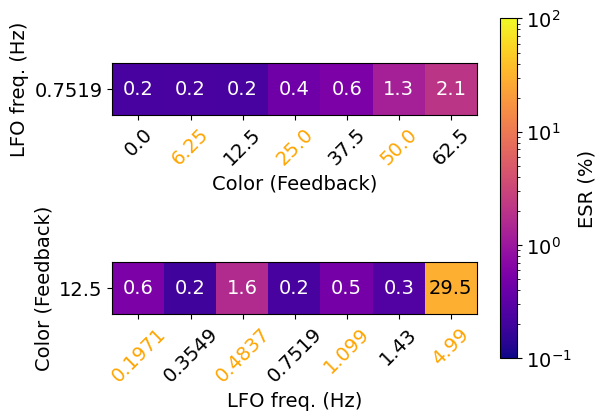}}
\caption{\itshape ESR(\%) results for model trained on the analog phaser under seen and unseen LFO frequencies and Color parameters. Seen cases are labeled black, and orange for unseen cases. Color is controlled over a fixed LFO frequency (Top). LFO frequency is controlled over a fixed Color rate (Bottom).}
\label{fig:analog_phaser_unseen}
\end{figure}

Figure \ref{fig:analog_phaser_unseen} presents the model performance on unseen control parameter settings.
The model shows effectiveness in predicting unseen feedback cases. However, it encounters challenges when inferring unseen LFO frequencies, especially those significantly higher than those observed within the seen frequency range. This difficulty may arise from complexities in modelling the nonlinear characteristics of analog circuits.

\subsubsection{Digital Flanger}


The flanger effect demonstrates higher spectral complexity compared to the phaser, introducing greater number of notches in the spectrogram of transfer function\cite{wright2020neural}. As a result, the flanger exhibits an overall higher ESR than the phaser.

\begin{figure}[t!]
\centerline{\includegraphics[width=\columnwidth]{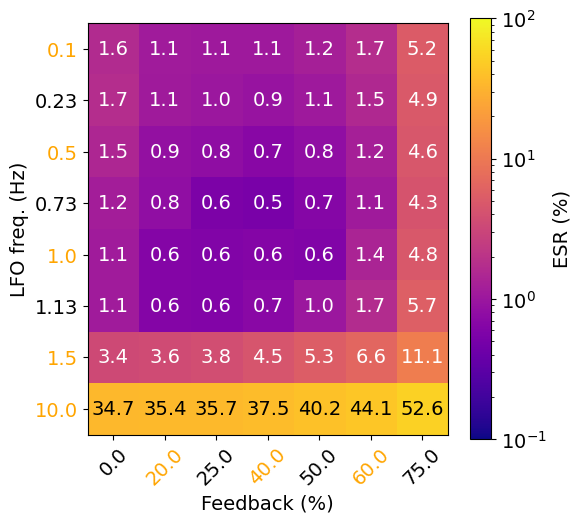}}
\caption{\itshape ESR(\%) results for model trained on digital flanger for under seen and unseen LFO frequencies and Feedback parameters. Seen cases are labeled black, and orange for unseen cases.}
\label{fig:digital_flanger}
\end{figure}

Figure \ref{fig:digital_flanger} displays the model's inference capability across both seen and unseen control parameters. The model exhibits strong performance for both seen and unseen feedback values. However, the model encounters difficulties in precisely predicting very high LFO frequencies, such as 10Hz. We anticipate that implementing adversarial loss \cite{kong2020hifi} could mitigate these issues for the extreme control parameter settings, and leave this as a consideration for future work.

\subsubsection{Analog Flanger}

\begin{figure}[t!]
\centerline{\includegraphics[width=7.5cm]{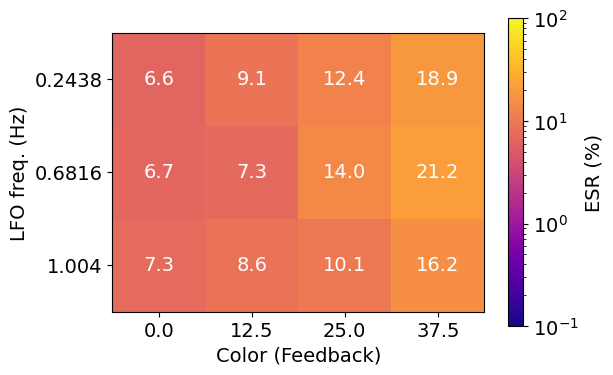}}
\caption{\itshape ESR(\%) results for model trained on the analog flanger under seen LFO frequencies and Color parameters.}
\label{analog_flanger_comparison}
\end{figure}

Figure \ref{analog_flanger_comparison} presents the ESR outcomes for the analog flanger model. Similar to previous experiments, the ESR decreases as feedback increases. Although the ESR is higher compared to the phaser scenario, it maintains an acceptable value overall.

\begin{figure}[h!]
\centerline{\includegraphics[width=7.5cm]{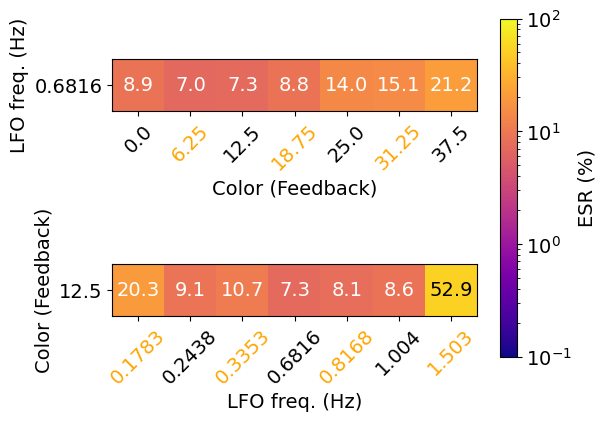}}
\caption{\itshape ESR(\%) results for model trained on the analog flanger under varying LFO frequencies and Color parameters. Seen cases are labeled black, and orange for unseen cases. Color is controlled over a fixed LFO frequency (Top). LFO frequency is controlled over a fixed Color rate (Bottom).}
\label{analog_flanger_unseen}
\end{figure}

Figure \ref{analog_flanger_unseen} illustrates the model's inference performance on unseen LFO frequency and feedback parameters. We observed that the model demonstrates robust interpolation for unseen feedback. Analogously, the model exhibited limitations in accurately inferring higher LFO frequencies.


\subsection{Robustness on long-term audio inference}

\begin{figure}[h]
\centerline{\includegraphics[width=7.5cm]{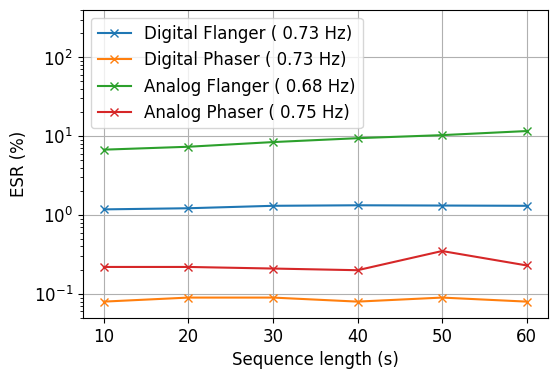}}
\caption{\itshape ESR(\%) for various input sequence length.}
\label{fig:long_seq}
\end{figure}

Robust long-term audio generation is essential for effect pedals, considering their use cases. Therefore, we analyzed our model's performance in long-term generation, with the results illustrated in Figure \ref{fig:long_seq}. 
In the case of the analog flanger, a minor increase in ESR was observed as the sequence length extended.
However, in general, our model demonstrated the ability to generate stable sequences of up to 60 seconds. Due to the periodic nature of the LFO, the LSTM was able to perform inference effectively even as the sequence length increased.

\subsection{Effect of training the model on multiple LFO frequencies}

\begin{figure}[]
\centerline{\includegraphics[width=\columnwidth]{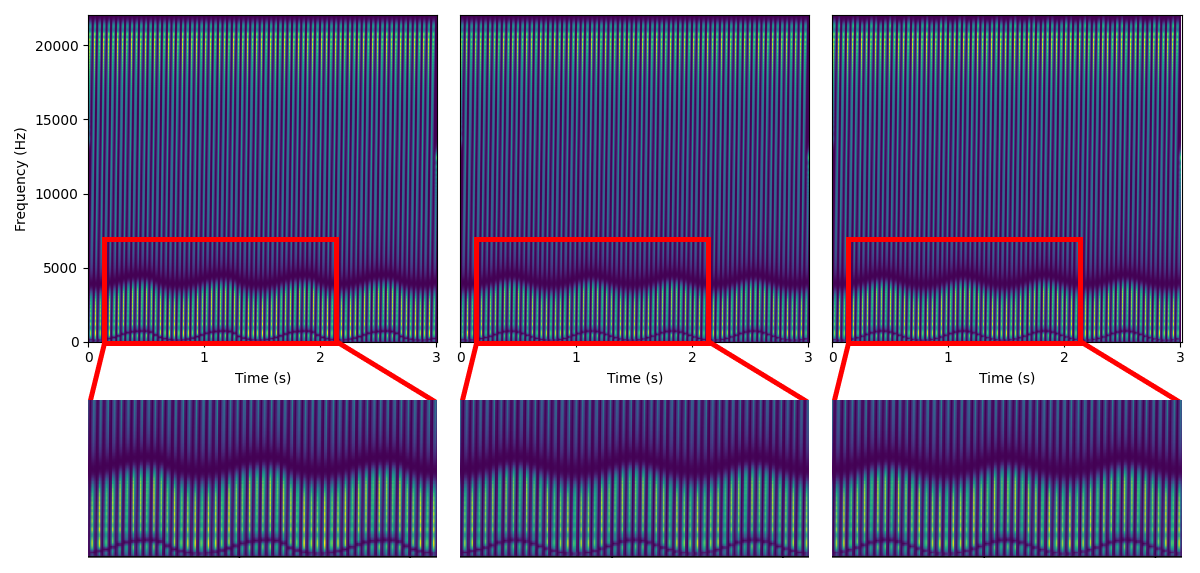}}
\caption{\itshape Analysis on the effect of training the model on multiple LFO frequencies. Each spectrogram illustrates the output of single LFO frequency model(left), multiple LFO frequency model(center), and target audio(right).}
\label{fig:multi-osc}
\end{figure}

To investigate the effects of training the model with multiple LFO frequencies, we compared the unseen LFO frequency inference capabilities of model trained with single LFO frequency and multiple LFO frequencies. As shown in Figure \ref{fig:multi-osc}, the spectrogram of the audio output from the model trained with a single LFO frequency (left) exhibits an inaccurate rightward-shifted notch. In contrast, our proposed architecture leveraging multiple LFO frequencies (center) displays a notch shape identical to the target audio (right). 
This outcome aligns with expectations, due to the restricted diversity of LFO phases encountered when the model is trained using a single oscillator.

\begin{figure*}[t!]
\center
\includegraphics[width=6.8in]{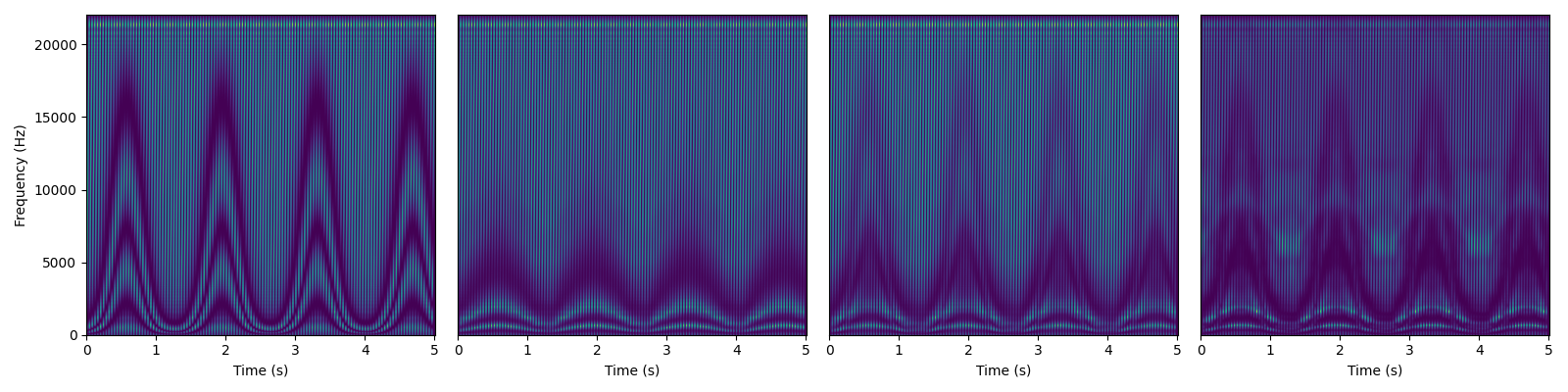}
\caption{\label{fig:multi_phaser}\itshape Analysis of the multiple-phaser effects trained results, depicted in spectrograms. For better visualization, we used chirp signal as input. The model is simultaneously trained with two digital phaser effects: Studio One Phaser$^2$ and Pedalboard phaser. Each spectrogram illustrates the model's output with the Pedalboard phaser embedding(leftmost), the model's output with the Studio One Phaser$^2$ embedding(left), a naive linear sum of the outputs from each embedding(right), and model's output with the interpolated embedding(rightmost).}
\end{figure*}

\subsection{Training on multiple phaser effects}\label{sec:multi_phaser}
To investigate the capability of our black-box modelling approach, we additionally trained our model concurrently on two distinct digital phasers: Studio One Phaser$^2$ and the Pedalboard Phaser\footnote{https://github.com/spotify/pedalboard}. The detailed control parameter settings for the training are described in Table \ref{tab:multiple_phaser_training}. The 2-dimensional embedding vector $c_{emb}$ was concatenated with $c_{fb}$ to be an input condition of the FiLM block. After training the model,
we derived a unique embedding vector interpolated between the two vectors each representing individual effects and analyzed the model's output upon this condition.
As illustrated in Figure \ref{fig:multi_phaser}, the model demonstrates consistently robust performance when conditioned with the learned embeddings for both phaser effects (leftmost and left). Notably, when the interpolated embedding is employed as the model condition (rightmost), the output exhibits a distinctive characteristic that lies between the two effects. For comparative analysis, we also present a result of a naive approach: a spectrogram of linearly summed audio from the two phaser effects (right).

\begin{table}[ht!]
\caption{\itshape Control parameter experiment settings for training multiple phaser effects.}
\label{tab:multiple_phaser_training}
\centering
\begin{tabular}{cc}
\Xhline{1.5pt}
\multicolumn{2}{c}{\textbf{Studio One Digital Phaser (Phaser$^2$)}} \\
\hline
\multicolumn{1}{c|}{Stages} & 6 \\
\multicolumn{1}{c|}{Speed (Hz)} & 0.37, 1.13 \\
\multicolumn{1}{c|}{Center} & 1300 \\
\multicolumn{1}{c|}{Range} & 80\% \\
\multicolumn{1}{c|}{Feedback (\%)} & 0, 25, 50, 75 \\
\multicolumn{1}{c|}{Stereo Spread} & 2.0\% \\
\multicolumn{1}{c|}{Mix} & 70.0\%  \\ 
\Xhline{1.5pt}
\multicolumn{2}{c}{\textbf{Pedalboard Phaser}} \\
\hline
\multicolumn{1}{c|}{Rate (Hz)} & 0.37, 1.13 \\
\multicolumn{1}{c|}{Centre Frequency (Hz)} & 1300 \\
\multicolumn{1}{c|}{Depth} & 0.5 \\ 
\multicolumn{1}{c|}{Feedback} & 0.0, 0.25, 0.5, 0.75 \\
\multicolumn{1}{c|}{Mix} & 0.5 \\
\hline
\end{tabular}
\end{table}

\section{Conclusion}

In this study, we presented a black-box model for LFO-driven effects targeting phaser and flanger. Our model offers controllability over LFO frequency and feedback parameters, by employing a combination of LSTM, MLP, and FiLM blocks, along with a unique training procedure optimizing multiple LFO frequencies. Comparative analysis with a baseline model revealed our model's superiority in both seen and unseen control parameters when modelling the phaser effect. Additionally, we found that our model can effectively generate audible flanger effects with reasonable controllability. Also, our model is capable of generating long sequences consistently with negligible amount of decrease in accuracy. Finally, we demonstrated our model's capability to train on two distinct phaser effects concurrently, resulting in creative mixed-up outputs between the effects.\\
However, our model faced challenges in inferring LFO frequencies that were outside the training data range, particularly evident in analog effect modelling. The performance was relatively lower for the flanger effect, due to its higher spectral complexity. Addressing these challenges may involve implementing methods capable of modelling complex distributions, such as adversarial loss. We aim to explore these solutions in future work.


\bibliographystyle{IEEEbib}
\bibliography{DAFx24_CONMOD}

\end{document}